# Two-Stage Self-Supervised Cycle-Consistency Network for Reconstruction of Thin-Slice MR Images

Zhiyang Lu[1], Zheng Li[1], Jun Wang[1, 2], Jun Shi[1, 2(✉)], Dinggang Shen[3, 4(✉)]

[1] Key laboratory of Specialty Fiber Optics and Optical Access Networks, Joint International Research Laboratory of Specialty Fiber Optics and Advanced Communication, School of Communication and Information Engineering, Shanghai University, China
[2] Shanghai Institute for Advanced Communication and Data Science, School of Communication and Information Engineering, Shanghai University, China
`junshi@shu.edu.cn`
[3] School of Biomedical Engineering, ShanghaiTech University, Shanghai, China
[4] Shanghai United Imaging Intelligence Co., Ltd., Shanghai 200230, China
`Dinggang.Shen@gmail.com`

**Abstract.** The thick-slice magnetic resonance (MR) images are often structurally blurred in coronal and sagittal views, which causes harm to diagnosis and image post-processing. Deep learning (DL) has shown great potential to reconstruct the high-resolution (HR) thin-slice MR images from those low-resolution (LR) cases, which we refer to as the slice interpolation task in this work. However, since it is generally difficult to sample abundant paired LR-HR MR images, the classical fully supervised DL-based models cannot be effectively trained to get robust performance. To this end, we propose a novel Two-stage Self-supervised Cycle-consistency Network (TSCNet) for MR slice interpolation, in which a two-stage self-supervised learning (SSL) strategy is developed for unsupervised DL network training. The paired LR-HR images are synthesized along the sagittal and coronal directions of input LR images for network pretraining in the first-stage SSL, and then a cyclic interpolation procedure based on triplet axial slices is designed in the second-stage SSL for further refinement. More training samples with rich contexts along all directions are exploited as guidance to guarantee the improved interpolation performance. Moreover, a new cycle-consistency constraint is proposed to supervise this cyclic procedure, which encourages the network to reconstruct more realistic HR images. The experimental results on a real MRI dataset indicate that TSCNet achieves superior performance over the conventional and other SSL-based algorithms, and obtains competitive qualitative and quantitative results compared with the fully supervised algorithm.

## 1 Introduction

Magnetic resonance imaging (MRI) is widely used for diagnosis of various diseases. Due to the issues of scanning time and signal-noise ratio [1], the thick-slice magnetic resonance (MR) images are generally acquired in clinical routine, which have low between-slice resolution along the axial direction [2]. Since the MR images with high spatial resolution are desirable to provide detailed visual information and facilitate further image post-processing [3], it is important to reconstruct high-resolution (HR) thin-slice MR images from these low-resolution (LR) thick-slice cases, which we refer to as the slice interpolation task in this work.

Deep learning (DL) has gained great reputation for image super-resolution (SR) in recent years [4-6]. Some pioneering works have also applied DL to reconstruct thin-slice MR image [7-10], which typically build a convolutional neural network (CNN) based model to perform slice interpolation with paired LR and HR images. However, due to the difficulty of collecting HR MR images in clinical practice, there are generally few datasets with sufficient paired samples for model training. Consequently, the performance of these DL-based slice interpolation models degrades seriously under the classical full voxel-to-voxel supervision.

Self-supervised learning (SSL), a commonly used approach for unsupervised learning, is one plausible solution to the lack of HR images, which can generate paired data only based on input data and enables network training without ground truth [11]. SSL has been successfully applied to various image reconstruction tasks, including image SR and video frame interpolation [12-14]. Motivated by the SSL-based SR strategy on 2D images, Zhao et al. proposed to downsample the axial slices of input LR images to form the LR-HR slice pairs for training an SR network, which was then applied to improve slice resolutions along the other two directions [15-17]. However, this algorithm ignores the contextual information between the axial slices during interpolation, which causes anatomical distortion in reconstructed axial slices. On the other hand, although the SSL-based video frame interpolation methods can be directly applied to MR slice interpolation by taking the consecutive axial slices as video frames [14], they cannot exploit the information from the dense slices along other directions to enhance interpolation performance. Therefore, the existing SSL-based approaches still cannot achieve improved reconstruction for slice interpolation in MRI.

To this end, we propose a novel Two-stage Self-supervised Cycle-consistency Network (TSCNet), in which a new two-stage SSL strategy is developed to train an interpolation network in an unsupervised manner to perform MRI slice interpolation. The paired LR-HR images are synthesized from the sagittal and coronal slices of input LR images for preliminary network training in the first-stage SSL. Then, since the central slice of the input triplet axial slices can be estimated from the two intermediate slices interpolated between each two adjacent slices of the triplet, a new cycle-consistency constraint is developed to supervise this cyclic interpolation procedure in the second-stage SSL, which can further refine the network and boost the interpolation performance. The experimental results on a real MRI dataset indicate its qualitatively and quantitatively comparable performance with the fully supervised algorithm.

The main contributions of this work are two-fold as follows:
1) A novel TSCNet is proposed to learn MRI slice interpolation with a two-stage SSL strategy only using LR images. The paired LR-HR images are synthesized based on sagittal and coronal directions for pretraining in the first-stage SSL, whereas the sparse axial slices can be exploited in the triplet manner as useful guidance for network refinement in the second-stage SSL. Training data with rich contexts about anatomical structures from all of the three directions can thus be exploited by the two-stage SSL to guarantee improved interpolation performance.

2) A new cycle-consistency constraint strategy is developed to supervise the cyclic interpolation procedure based on the input triplet axial slices in the second-stage SSL to further refine the pretrained interpolation network. This strategy allows for SSL free from unwanted axial downsampling and can effectively encourage the network to interpolate more realistic intermediate slices required in HR images.

## 2   Method

Fig. 1 illustrates the framework of the proposed TSCNet. An interpolation network trained by the strategy of two-stage SSL will interpolate the intermediate slices between the input two adjacent slices of thick-slice LR images. The general pipeline of the two-stage SSL is described as follows:

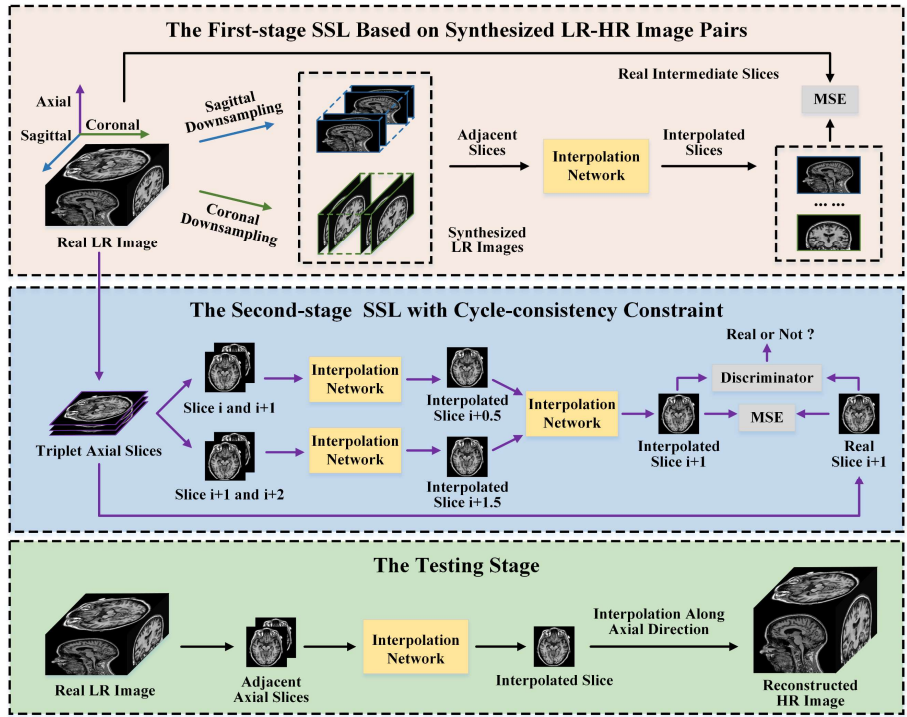

**Fig. 1.** Framework of the Two-stage Self-supervised Cycle-consistency Network. The slice $i$+0.5 denotes the interpolated slices between the $i$-th and ($i$+1)-th axial slices of the LR image.

1) In the first-stage SSL, the input LR image is downsampled into synthesized LR-HR data based on the dense slices along its sagittal and coronal directions. Sufficient paired samples can thus be generated for pretraining the interpolation network to learn the rich knowledge about anatomical structures

2) In the second-stage SSL, the central slice of input triplet axial slices is estimated in a cyclic interpolation procedure for network refinement, which is supervised by the powerful cycle-consistency constraint to implicitly force the interpolated intermediate slices needed in reconstructed HR images to be more realistic.

3) In the testing stage, the well-trained interpolation network interpolates the slices along the axial direction of input LR images to reconstruct HR thin-slice images.

### 2.1 Interpolation Network

TSCNet provides an effective two-stage SSL strategy for training the interpolation network with any structure. Specifically, the network structure used in this work is shown in Fig. 2. Two paralleled feature extraction modules are utilized to extract high-level features respectively from the input adjacent slices of LR images. The generated two sets of features that contain rich contextual information are combined by concatenation into joint feature representations, which are then refined through a slice reconstruction module to reconstruct the intermediate slice.

The residual dense network (RDN) [18] is utilized to construct each of the three modules in the interpolation network. Specifically, the RDN contains cascaded residual dense blocks (RDB), which is a powerful convolutional block that takes advantage of both residual and dense connections to fully aggregate hierarchical features, the cascaded RDB can thus provide sufficient capability of feature extraction and refinement, which enables the interpolation network to get improved reconstruction performance. More details about the structure of RDN can be referred to [18].

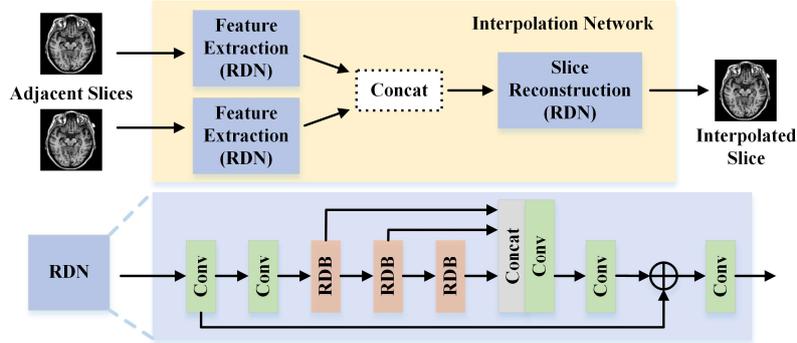

**Fig. 2.** Architecture of the interpolation network and the residual dense network (RDN).

### 2.2 The First-stage SSL Based on Synthesized LR-HR Image Pairs

Due to the absence of real HR images, SSL requires to synthesize the paired LR-HR training data from the real LR images to train the interpolation network. One plausible method of SSL is to downsample the LR images along the axial directions. However, since the axial slices are originally sparse, this strategy can only get the paired data that have severe discrepancy between the input adjacent slices, which easily misguides the network for coarse interpolation. We thus synthesize LR-HR pairs by downsampling LR images on the dense slices along the sagittal and coronal directions for network pretraining in the first-stage SSL.

Given a real LR image $I \in \mathbb{R}^{X \times Y \times Z}$, we downsample the image by factor 2 along its sagittal and coronal directions, respectively, to get two synthesized LR images $I_{\downarrow sag} \in \mathbb{R}^{(X/2) \times Y \times Z}$ and $I_{\downarrow cor} \in \mathbb{R}^{X \times (Y/2) \times Z}$, and regard $I$ as the HR version of these two images. Subsequently, the adjacent two sagittal slices in $I_{\downarrow sag}$ are extracted as the input data of training, and the corresponding real intermediate slices are derived from $I$ as the ground truth, constructing a training dataset $\{S_{\downarrow sag}^i, S_{\downarrow sag}^{i+1}, S_{sag}^{2i}\}_{i=1}^{(X/2)-1}$, in which $S_{\downarrow sag}^i$ denotes the $i$-th sagittal slice in $I_{\downarrow sag}$ and $S_{sag}^{2i}$ is the $2i$-th sagittal slice in $I$. With the same operation, we can also get $\{S_{\downarrow cor}^i, S_{\downarrow cor}^{i+1}, S_{cor}^{2i}\}_{i=1}^{(Y/2)-1}$ as training data based on the coronal slices of $I_{\downarrow cor}$ and $I$.

When two adjacent slices of the synthesized LR image are fed into the interpolation network, the missing intermediate slice can be generated. To train this model, we utilize the MSE loss function to enforce the pixel-wise consistency between the interpolated slice and the real intermediate slice, which can be formulated as:

$$l_{mse} = \frac{1}{X-2}\sum_{i=1}^{(X/2)-1}\left\|S_{sag}^{2i} - \mathcal{F}(S_{\downarrow sag}^i, S_{\downarrow sag}^{i+1})\right\|^2 + \frac{1}{Y-2}\sum_{i=1}^{(Y/2)-1}\left\|S_{cor}^{2i} - \mathcal{F}(S_{\downarrow cor}^i, S_{\downarrow cor}^{i+1})\right\|^2 \quad (1)$$

where $\mathcal{F}(\cdot)$ denotes the mapping function of the interpolation network.

### 2.3 The Second-stage SSL with Cycle-consistency Constraint

The contexts of coronal and sagittal slices in the original LR images have been exploited within the synthesized dataset to guide the training of interpolation network in the first-stage SSL. Thus, the pretrained interpolation network can already perform coarse slice interpolation. However, the contexts along the axial direction are still ignored in training, and the used MSE loss function easily introduces image smoothness and artifacts, resulting in the reconstruction of unrealistic axial slices. We thus design a cyclic interpolation procedure using triplet axial slices and develops a cycle-consistence constraint to supervise this procedure, which further refine the interpolation network to achieve more realistic interpolation results in the second-stage SSL.

Given a real LR image $I \in \mathbb{R}^{X \times Y \times Z}$, a set of consecutive triplet axial slices can be extracted from $I$ as $\{S_{axi}^i, S_{axi}^{i+1}, S_{axi}^{i+2}\}_{i=1}^{Z-2}$, in which $S_{axi}^i$ denotes its $i$-th axial slice. Then, the two sets of adjacent two slices in a triplet are fed into the interpolation network, respectively, to interpolate the intermediate slices, which can be formulated as:

$$\bar{S}_{axi}^{i+0.5} = \mathcal{F}(S_{axi}^i, S_{axi}^{i+1}) \quad (2)$$

$$\bar{S}_{axi}^{i+1.5} = \mathcal{F}(S_{axi}^{i+1}, S_{axi}^{i+2}) \quad (3)$$

in which $\bar{S}_{axi}^{i+0.5}$ denotes the interpolated slice between $S_{axi}^i$ and $S_{axi}^{i+1}$, whereas $\bar{S}_{axi}^{i+1.5}$ is the interpolated slice between $S_{axi}^{i+1}$ and $S_{axi}^{i+2}$.

Given $\bar{S}_{axi}^{i+0.5}$ and $\bar{S}_{axi}^{i+1.5}$, the central slice $S_{axi}^{i+1}$ of the original input triplet can be reversely estimated through the interpolation network, which is formulated as:

$$\bar{S}_{axi}^{i+1} = \mathcal{F}(\bar{S}_{axi}^{i+0.5}, \bar{S}_{axi}^{i+1.5}) \tag{4}$$

We can thus adopt $S_{axi}^{i+1}$ as the ground truth to supervise this cyclic procedure through a cycle-consistency constraint, which implicitly enforces the model to reconstruct realistic $\bar{S}_{axi}^{i+0.5}$ and $\bar{S}_{axi}^{i+1.5}$. Specifically, in addition to the MSE loss function, we add an adversarial loss to constrain the similarity between $S_{axi}^{i+1}$ and $\bar{S}_{axi}^{i+1}$, which aims to overcome the issue of image smoothness in reconstructed slices.

To this end, we adopt the discriminator in SRGAN to distinguish between the estimated $\bar{S}_{axi}^{i+1}$ and the real $S_{axi}^{i+1}$, whose objective function is defined as follows [19]:

$$l_D = \frac{1}{Z-2} \sum_{i=1}^{Z-2} [-\log D(S_{axi}^{i+1}) - \log(1 - \bar{S}_{axi}^{i+1})] \tag{5}$$

where $D(\cdot)$ denotes the function of the discriminator.

Meanwhile, the cycle-consistency constraint is formulated as:

$$l_{cyc} = \frac{1}{2Z-4} \sum_{i=1}^{Z-2} \|S_{axi}^{i+1} - \bar{S}_{axi}^{i+1}\|^2 - \frac{\lambda}{Z-2} \sum_{i=1}^{Z-2} [\log(\bar{S}_{axi}^{i+1})] \tag{6}$$

where $\lambda$ is the parameter to balance the MSE and adversarial loss.

Finally, we utilize $l_{mse}$ together with $l_{cyc}$ to guarantee the stability of training, and the total loss of the second-stage SSL for network refinement is defined as:

$$L = l_{mse} + l_{cyc} \tag{7}$$

## 3 Experiments

### 3.1 Dataset

The proposed TSCNet algorithm was evaluated on 64 T1 MR brain images selected from the publicly available Alzheimer's Disease Neuroimaging Initiative (ADNI) dataset [20]. All of the MR images were sampled at 1×1×1 mm3 and zero-padded to the voxel size of 256×256×256 as the real HR thin-slice images. Then we downsampled the isotropic volumes by factors of 2 along the axial direction to generate the LR thick-slice images with the voxel size of 256×256×128.

### 3.2 Experimental Design

To evaluate the performance of our proposed TSCNet algorithm, we compared it with the following algorithms for slice interpolation:
1) Trilinear Interpolation: A conventional algorithm to interpolate 3D images.
2) EDSSR [15]: An SSL-based DL algorithm for MR slice interpolation. It generated paired LR-HR axial slices from LR images to train an SR network, which was then used to perform SR on coronal and sagittal slices to reconstruct HR images with the technique of Fourier Burst Accumulation. It is worth mentioning that a data augmentation operation was applied on EDSSR to form the SMORE (3D) algorithm proposed in the more recent paper [17]. However, we did not conduct any similar operations in our experiments.

3) Full-supervised Interpolation Network (FSIN): A fully supervised algorithm based on the interpolation network, which used the adjacent axial slices of real LR images as input and the corresponding intermediate slices in real HR images as ground truth, and is trained with the common MSE loss function.

An ablation experiment was also performed to compare our proposed TSCNet algorithm with the following two variants of TSCNet:

1) TSCNet with Adapted First-stage SSL (TSCNet with AFS): This algorithm modified the first-stage SSL based on the common methods for video frame interpolation to evaluate the effectiveness of this stage, which downsampled real LR images only along the axial direction to synthesize paired LR-HR training samples.
2) TSCNet without Second-stage SSL (TSCNet w/o SS): This algorithm only adopted the first-stage SSL to train the interpolation network, and removed the second-stage SSL to validate the effectiveness of proposed cycle-consistency constraint.

We performed the 4-fold cross-validation to evaluate the performance of different algorithms. The commonly used peak signal-to-noise ratio (PSNR) and structural similarity index (SSIM) were adopted as the evaluation indices. The results of all folds were reported with the format of the mean ± SD (standard deviation).

## 3.3 Implementation Details

In our implementations, all real images were normalized to voxel values between 0 and 1 as the network inputs. Each module in TSCNet removed the upsampling operation in original RDN and has three RDB with five layers. For training TSCNet, the first-stage strategy was conduct for 100 epochs, and the discriminator and the interpolation network were optimized alternately for another 150 epochs in the second-stage, and the parameter $\lambda$ in $l_{cyc}$ was set to 0.1. The batch size was set to 6 for the two adjacent coronal or sagittal slices, and 3 for triplet axial slices. All of the DL-based algorithms were trained by an Adam optimizer with a learning rate of 0.0001.

## 3.4 Experimental Results

Fig. 3 shows the interpolated slices along the axial directions in the HR thin-slice images reconstructed by different slice interpolation algorithms. It can be found that our proposed TSCNet can reconstruct more realistic slices with reference to the ground truth compared with the conventional and existing SSL-based algorithms. Trilinear interpolation generally reconstructs blurred results, and EDSSR generates inaccurate tissue structures. TSCNet also achieves superior performance over its two variants with clearer tissue boundaries and less artifacts in ablation study, which validates the effectiveness of each stage in the two-stage SSL.

Moreover, TSCNet can get visually comparable performance with the fully-supervised FSIN, and even better results in the third and fourth rows in Fig. 3, which demonstrates its robust capability for interpolating realistic slices.

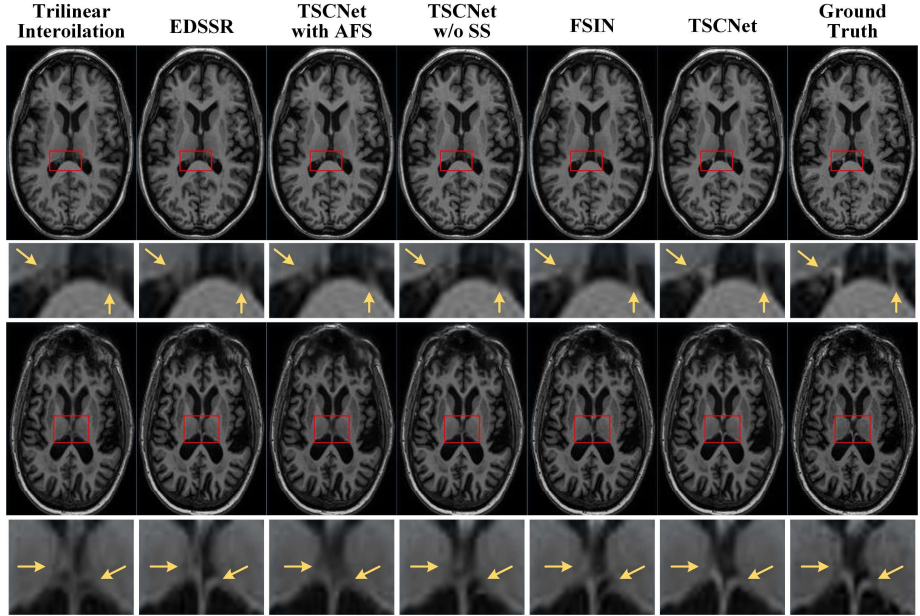

**Fig. 3.** Visual comparison of different slice interpolation algorithms on ADNI dataset.

Table 1 gives the qualitative results of experiments, which indicate that TSCNet outperforms the conventional trilinear interpolation and the SSL-based EDSSR with the PSNR of 36.14 ± 0.43 and SSIM of 0.9419 ± 0.0040. Specifically, TSCNet improves by 0.7 dB and 0.0075 on the two indices compared with EDSSR, which demonstrates that more contextual information related to the missing intermediate slices can be learned through the proposed two-stage SSL strategy for superior performance. TSCNet also gets an improvement by at least 0.22dB and 0.0025 on PSNR and SSIM over its two variants in ablation study, which validates the effectiveness of each stage in proposed SSL strategy. It can be concluded that more training samples can be obtained in TSCNet to enhance the robustness of interpolation network compared with TSCNet with AFS, whereas the cycle-consistency constraint utilized in the second-stage of TSCNet can benefit to reconstruct more realistic HR images.

It is worth mentioning that TSCNet achieves close quantitative performance to the fully-supervised algorithm. Besides, it takes advantage of SSL to avoid the requirement for the expensive HR ground truth, and can even reconstruct visually better HR images as shown in the qualitative results.

**Table 1.** Quantitative results of different algorithms for slice interpolation.

| Algorithm | Category | PSNR (dB) | SSIM |
| --- | --- | --- | --- |
| Trilinear Interpolation | Convention | 34.03 ± 0.37 | 0.9139 ± 0.0037 |
| EDSSR | SSL | 35.44 ± 0.18 | 0.9344 ± 0.0012 |
| TSCNet with AFS | SSL | 35.86 ± 0.43 | 0.9364 ± 0.0036 |
| TSCNet w/o SS | SSL | 35.92 ± 0.44 | 0.9394 ± 0.0043 |
| TSCNet (ours) | SSL | 36.14 ± 0.43 | 0.9419 ± 0.0040 |
| FSIN | Full Supervision | 36.53 ± 0.39 | 0.9447 ± 0.0039 |

## 4     Conclusion

In summary, a novel TSCNet algorithm is proposed to address the slice interpolation problem for reconstruction of HR thin-slice MR images. It develops an SSL-based strategy with a novel cycle-consistency constraint to exploit contextual information along all directions for training the interpolation network only based on LR thick-slice input, which enables improved reconstruction free from the guidance of the clinically rare HR ground truth. The experimental results validate the effectiveness of TSCNet by demonstrating its superior performance over the traditional and other SSL-based algorithms, and comparable performance with fully supervised algorithms.

In future work, TSCNet will be combined with the fully supervised algorithm to further enhance the reconstruction performance. The novel structure of the interpolation network will also be explored to boost the interpolation performance in TSCNet.

**Acknowledgements.** This work is supported by National Natural Science Foundation of China (81830058) and the 111 Project (D20031).

10